\documentstyle[11pt,aaspp4,tighten]{article}
\received{October 7th, 1998}
\accepted{January 12th, 1999}
\def\today{\rightline{\ifcase\month\or
        January\or February\or March\or April\or May\or June\or
        July\or August\or September\or October\or November\or December\fi
        \space\number\day, \number\year}}
\def\etal{{\it et al.}\ }

\def\s-1{{\rm\,s^{-1}}}

\def\spose#1{\hbox to 0pt{#1\hss}}
\def\C3H2{{\rm\,\rm C_3H_2}}
\def\NH3{{\rm\,\rm NH_3}}
\def\HOCO+{{\rm\,\rm HOCO^+}}
\def\lta{\mathrel{\spose{\lower 3pt\hbox{$\mathchar"218$}}
     \raise 2.0pt\hbox{$\mathchar"13C$}}}
\def\gta{\mathrel{\spose{\lower 3pt\hbox{$\mathchar"218$}}
     \raise 2.0pt\hbox{$\mathchar"13E$}}}

\begin{document}

\font\twelvei = cmmi10 scaled\magstep1 
       \font\teni = cmmi10 \font\seveni = cmmi7
\font\mbf = cmmib10 scaled\magstep1
       \font\mbfs = cmmib10 \font\mbfss = cmmib10 scaled 833
\font\msybf = cmbsy10 scaled\magstep1
       \font\msybfs = cmbsy10 \font\msybfss = cmbsy10 scaled 833
\textfont1 = \twelvei
       \scriptfont1 = \twelvei \scriptscriptfont1 = \teni
       \def\mit{\fam1 }
\textfont9 = \mbf
       \scriptfont9 = \mbfs \scriptscriptfont9 = \mbfss
       \def\bmit{\fam9 }
\textfont10 = \msybf
       \scriptfont10 = \msybfs \scriptscriptfont10 = \msybfss
       \def\bmsy{\fam10 }

\def\etal{{\it et al.~}}
\def\eg{{\it e.g.}}
\def\ie{{\it i.e.}}
\def\lsim{\raise0.3ex\hbox{$<$}\kern-0.75em{\lower0.65ex\hbox{$\sim$}}} 
\def\gsim{\raise0.3ex\hbox{$>$}\kern-0.75em{\lower0.65ex\hbox{$\sim$}}} 
\pagestyle{plain}
%\today
\title{A CATALOGUE OF OPTICALLY SELECTED CORES \footnote[1]
  {to appear in the Astrophysical Journal Supplement Series}}

\author{Chang Won Lee \& Philip C. Myers}
\vskip 0.2in
\affil{Harvard-Smithsonian Center for Astrophysics,}
\affil {60 Garden Street, MS 42, Cambridge, MA  02138, USA}
 
\affil{E-mail: cwlee@cfa.harvard.edu, pmyers@cfa.harvard.edu}
\vskip 1in
\begin{abstract}
We present  a new catalogue of 406 dense cores optically selected 
by using the STScI Digitized Sky Survey (DSS).
In this catalogue 306 cores have neither an Embedded YSO (EYSO) nor 
a Pre-Main-Sequence (PMS) star, 94 cores have EYSOs (1 core has both an EYSO and a PMS star), 
and 6 cores have PMS star only.
Our sample of dense cores in the catalogue is fairly complete within
a category of northern Lynds class 5, 6 clouds, and southern Hartley et al. (1986)'s class A clouds,
providing a database useful for the systematic study of dense cores.
Most of the cores listed in the catalogue have diameters between $0.05 - 0.36$ pc with 
a mean of $\sim 0.24$ pc.
The sizes ($\sim 0.33$ pc in the mean ) of cores with EYSOs are found to be usually larger than 
the sizes  ($\sim 0.22 $ pc in the mean) of starless cores. The typical mean gas density 
of the cores is $\sim7\times 10^3$ cm$^{-3}$.
Most of the cores are more likely elongated than spherical (mean aspect ratio: $\sim 2.4$).
The ratio of the number of cores with EYSOs to the number of starless cores
for our sample  is about 0.3, suggesting that the typical lifetime of starless cores 
is $0.3-1.6$ Myr, about 3 times longer than the duration of the Class 0 and Class I phases.
This lifetime is shorter than expected from models of ambipolar diffusion, by 
factors of $2-44$.
\end{abstract}


\clearpage

\section{INTRODUCTION}

Low mass stars are generally believed to form through the inward gravitational collapse
of ``dense cores'' with density $\ga$ a few $10^4$ cm$^{-3}$ which are expected 
to form via ambipolar diffusion processes
(Shu, Adams \& Lizano 1987; Ciolek \& Mouschovias 1995).
However the knowledge of such simple processes in star
formation is almost entirely theoretical because there is still very little direct 
observational support. It is therefore desirable to conduct unbiased systematic surveys of dense cores
to search for evidence of core evolution and star-forming motion.

The optical dark cloud catalogues of Lynds (1962), Feitzinger \& Stuwe (1984), and
Hartley et al. (1986, hereafter HMSTG), based on visual inspection of optical sky survey prints,
have provided us the most basic and complete set of sources for 
the study of star-forming regions, 
although these catalogues themselves could not  give accurate positions for 
a systematic survey of 
dense cores because the positions of dark regions were determined approximately by eye.
Guidance of these catalogues initiated to make more accurate and detailed catalogues.
For the northern hemisphere, 
Myers, Linke, \& Benson (1983, hereafter MLB) have collected 90 small (angular size $<~5'$) 
opaque spots through visual inspection of the Palomar Sky Atlas prints
to conduct a systematic survey in $^{13}$CO, C$^{18}$O (1-0) lines , and later
NH$_3$ (Benson \& Myers 1989, hereafter BM). Clemens \& Barvainis (1988, hereafter CB) have also identified 
248 molecular clouds with optical sizes smaller than $10'$ by similar procedures
to MLB's method.
For the southern hemisphere, Vilas-Boas et al. (1994) selected 101 condensations 
with average optical size less than $7'$ and visual extinction greater than $2.5^m$ 
from ESO J plates, visual extinction maps (Hetem et al. 1988), 
and the HMSTG catalogue to observe them in $\rm ^{13}CO$ and $\rm C^{18}O$
$(1-0)$ lines.  
Bourke et al. (1995a) have compiled 169 Bok-globule-like small ($<10'$) dark clouds 
from visual inspection
of the SERC Schmidt survey J plates primarily with the guidance of the HMSTG catalogue
and observed  all the cores  in ammonia (1,1) line (Bourke et al. 1995b). 

These efforts resulted in an excellent database for the observational study of low mass star formation.  
However, the samples given in these catalogues were limited to a few regions 
and many cores are still missing even toward the searched area,
so these previous studies are not complete.
Moreover, most opaque positions of the dark clouds  in these catalogues were determined  by eye, 
so that positions are not only very subjective, but also not accurate
enough for an efficient survey in molecular lines.

Now, by the efforts of the Space Telescope Science Institute (STScI), the Digitized Sky 
Survey (DSS) data are available for us.
These data allow us to computerize the selection of cores and 
accurately determine their  positions. 
With the guidance of the Lynds (1962) catalogue for the northern sky and the Hartley et al. 
(1986) catalogue for the southern sky, in addition to several reference catalogues 
that have been used for numerous 
molecular line studies (see in \S 2 for detailed references),   
we created a new catalogue of dense cores using the STScI DSS. 
Our catalogue presents the most complete list of optically selected dense cores whose positions 
were accurately measured within a few arc seconds.
The actual accuracy of positions is dependent only upon the procedures 
for smoothing the data and centering the intensity minima of the core. 

In the following section we describe in detail how we selected the cores. Then we analyze the statistics 
of the physical parameters of our sample and discuss the implications of core statistics 
for the present standard model
of isolated star formation. Finally we summarize our results in \S 3.
\section{THE CATALOGUE}

\subsection{Selection of Cores }

Our method for the selection of cores is to  image an area,  usually $0^\circ.5\times 0^\circ.5$
and occasionally $1^\circ.0\times 1^\circ.0$ 
around a reference position, and to make a contour map of optical extinction 
to select the multiple local minimum positions of intensity. 
For guiding reference positions, we used the following;
BM (NH$_3$ peaks), Onishi et al. (1996; C$^{18}$O peaks),
Codella et al. (1996; NH$_3$ peaks ), Lemme et al. (1996; NH$_3$ peaks), and
Lynds (1962; every Lynds clouds with opacity 5 and 6) for northern clouds, 
and  Vilas-Boas et al. (1994; C$^{18}$O peaks) and HMSTG 
(clouds with density class A) for southern clouds (Table 1).
Digitized data \footnote[1] 
{Images from the northern plates (POSS) were based on photographic data of the National
Geographic Society -- Palomar Geographic Society
to the California Institute of Technology, and images from the southern plates (SERC-J)
were based on photographic data obtained using the UK Schmidt Telescope.
The plates were processed into the present compressed
digital form with their permission. The UK Schmidt
Telescope was operated by the Royal Observatory Edinburgh, with funding from the UK Science
and Engineering Research Council, until 1988 June, and thereafter by the Anglo-Australian
Observatory.
The Digitized Sky Survey was produced at the Space
Telescope Science Institute under US Government grant NAG W-2166.}
were retrieved from the STScI DSS using the `getimage' software package provided by  STScI.
Image processing was conducted with the IRAF software package\footnote[2]{
IRAF is distributed by the National Optical Astronomy Observatories,
    which are operated by the Association of Universities for Research
    in Astronomy, Inc., under cooperative agreement with the National
    Science Foundation.}(Tody 1993).
An accurate  reading (within error of a few arc seconds) of coordinates of 
local intensity minima in the contour map overlapped
in the grey image in IRAF was made possible by browsing images in World Coordinate Space (Mink 1995)
with the ``SAOimage'' - X11 window based image display program (VanHilst 1990). 
However, the real accuracy of our positions was dependent upon the grid size we used to make 
smooth contours and center the intensity minima of the core, usually about $15''\sim 50''$.
In order to check the reliability of our method, we compared our new positions 
with peak positions of NH$_3$ maps (with beam resolution of $87''$) 
for 22 cores by BM.
The difference [$68''\pm 10''$ (mean$\pm$s.e.m.)] between two positions 
was found to be smaller than the beam resolution of the NH$_3$ maps, 
meaning  that our method
is fairly reliable for determining the intensity minima positions of new cores.  

Fig. 1 shows one example of new core selection using the dark cloud, L1622.
This example is typical in that several distinct intensity
minima are obtained for each reference position.
In total, we examined 250 reference positions (Table 1), and identified 406 local maxima of 
optical extinction (Table 2), which we call cores.
For all sources, we searched the IRAS Point Source
Catalogue using the `riras' package (Mink 1998) to determine which sources
have embedded IRAS point sources.
In the search, we assumed the IRAS point source to be an ``$Embedded$'' Young Stellar Object (EYSO) 
when it satisfies all of the following criteria;

(1) There should be detections in at least two of the four IRAS wavelength bands. 

(2) The detected fluxes ($F$) should be greater in the longer wavelength. 
However, IRAS point sources 
with $F_{100\micron} < F_{60\micron}$, or $F_{100\micron} <F_{25\micron} $, or 
$F_{100\micron} <F_{12\micron} $ were considered to be EYSOs 
as long as $F_{60\micron} > F_{25\micron} >F_{12\micron}$. 

(3) The projected position of the IRAS source on the sky should be enclosed by 
the contour of the least extinction level of core in the optical image.

With these criteria, we searched all cores to determine whether there are EYSOs 
within a box with a size of 2 times the geometric mean of the FWHM of a core.
The FWHM of the core  was obtained 
from an extinction contour map of the core as described in the next section. 
We found 94 sources with EYSOs, and 306 cores without EYSOs.
Our criteria are very similar to those of BM, but differ from 
those of Beichman et al. (1986) and Wood et al. (1994) in that ours were designed to
preferentially select embedded protostars (Class 0 and Class I YSOs) 
and therefore not to include Pre-Main-Sequence (PMS) star.
To find a possible PMS star we used color-color diagram criteria in 12, 25,
and 60 $\micron$ IRAS bands of Weintraub (1990) ; 
$-2.00 < {log(\nu_{12}F_{12}/\nu_{25}F_{25}) \over log(\nu_{12}/\nu_{25})} < 1.35$ and
$-1.75 < {log(\nu_{25}F_{25}/\nu_{60}F_{60})\over log(\nu_{25}/\nu_{60})} < 2.20$.
Five cores were found to have possible PMS stars.
We also searched the catalogue of PMS stars by Herbig \& Bell (1988) in which  
PMS stars were identified by optical spectroscopic data. 
From this search 2 more sources were found to include PMS stars within their projected extents.
Thus we believe that 7 sources may have PMS stars. 
We note that 1 of 7 contains EYSO as well. 

In summary, of 406 identified cores, 306 were found to have neither an EYSO nor a PMS star, 
94 were found to have EYSOs (1 core has both an EYSO and a PMS star), and 6  to have PMS star only. 
Henceforth we call the 306 cores with neither EYSO nor PMS star `starless cores'.
In the 9th column of the Table 2, we marked the starless cores as `N' and cores with EYSOs as `Y',
cores with PMS star as `PM' or especially `PM*' for the PMS star known by optical spectroscopy 
(Herbig \& Bell 1988), and cores with both EYSO and PMS star as `Y/PM'.
The IRAS point sources and the PMS stars,
believed to be associated with our optically selected cores are listed in Table 3.

\subsection{Statistics of Physical Parameters  of Cores}

In this section  we discuss the statistics of the physical parameters of cores.
Our sample statistics are believed to be representative of the core
properties since the cores were selected from Lynds class 5, 6, and HMSTG's class A, 
which, taken together, comprise a nearly complete set of the densest and the most opaque clouds in the sky.

We present several physical parameters of the optically selected cores in Table 2. 
The sizes of major axis (`a' in the column 5 of Table 2) and minor axis 
(`b' in the column 6 of Table 2), 
and position angle (`PA' in the column 8 of Table 2) of the FWHM of the cores 
were approximately measured by an ellipse template, ruler, and protractor.
The `$-$' and `$+$' in the position angle mean `clockwise' and `counter-clockwise'
with respect to Equatorial north, respectively.
The largest errors for these parameters are from the uncertainties in the selection
of the background level of the core which determines the size of the FWHM of the core, and 
in the eye fitting of the ellipse template to the core.  
The errors obtained by repeating measurements are usually within about 10$\%$ 
of the size in the size measurement, and within about 5 degrees in the PA measurement.
For the sources which are not elliptical in shape, 
their `equivalent' diameters (`R' in column 7 of Table 2) were measured. 
The equivalent diameter of a core is defined as $2\times 
({A\over\pi})^{0.5}$ where A is the area within the FWHM contour of the core.
Following are our results and discussion on the statistics of these physical parameters.

{\bf \it 2.2.1 The darkness contrast of cores}

In the 10th column of the Table 2 we give an indication of a Darkness Contrast (DC) of 
the dark core to the background in the DSS image.
This contrast was calculated according to $ {I_B - I_C} \over I_B$, where $I_B$ and $I_C$ are 
the data values (DN) by DSS [that are on a scale of photographic density ($\gamma$) in the
expression of $\gamma=DN/6553.4$ (Postman 1996)] for the background and 
the core, respectively. The $I_B$ and $I_C$ values were approximately determined by using a cut-profile 
of the data values crossing the core. The $I_B$ and $I_C$ would correspond to a constantly flat
level and  a minimum level of the profile, respectively. 
Fig. 2 shows an example of how we adopted $I_B$ and $I_C$ for L1622. This cut-profile is 
the one crossing two intensity minima of L1622A and L1622B.

According to this value of the DC, we classified cores into
four quartiles, 1 being the least dark to 4 the most dark so 
that each group contains $25\%$ of total number of cores.
It is noted that the Lynds 6 and HMSTG A clouds more likely consist of cores with a DC of 3 or 4,
 while Lynds 5 clouds more likely have cores with a DC of 1 or 2 (Fig. 3).
However we also note that if the dark cores are located at the dark background, the DC values 
tend to have a low value. This explains why  many Lynds 6 cores also have a DC value of 1 in Fig. 3. 

{\bf \it 2.2.2 The size and the aspect ratio of cores}

For evaluation of these properties, we do not consider cores with more than two
peaks in the contour map because of ambiguity in the size information of such cores, 
but do consider cores with only one peak in the map whose size and aspect ratio can be better 
determined.
The distribution of the apparent angular size of cores is shown with the geometric mean 
$(ab)^{0.5}$ of cores in Fig. 4, indicating that
most of the cores ($\sim$70 percent) have angular size between $1'\sim4'$ with a mean
($\pm$ s.e.m.) of $3.7'(\pm 0.1)$. 
This distribution is similar to that found by previous study of CB. 
The  size distribution of starless cores, which comprise a majority, resembles the  size distribution 
of all cores, while the size of cores with EYSOs is uniformly  distributed between $1'\sim 8'$.
The interesting thing is that the mean size $(5.1\pm0.4)'$ of the cores with EYSOs is  
larger than that $(3.3\pm0.1)'$ of starless cores.  

We also find a corresponding difference in the distribution of the linear sizes, 
as illustrated in Fig. 5.
Distances (in 12th column of Table 2) of cores for 
determination of the linear sizes have been obtained from
numerous papers (see captions in Table 2) referred  by Dame et al. (1987), BM, 
Vilas-Boas et al. (1994), Hilton \& Lahulla (1995), and Chen et al. (1997).
We adopted the distances given by those papers if our sources are in a vicinity (within 1 degree) of 
the clouds whose distances are known,  and/or LSR velocities 
(known from a recent CS (2-1) and N$_2$H$^+$ (1-0) survey; Lee et al. 1998) 
of our sources are within $\pm 2$ km s$^{-1}$ of those of the known  clouds. 
Linear sizes of more than 80 percent of all the cores are between $0.05 - 0.35$ pc and the mean size ($\pm$ s.e.m.)
of the cores is about 0.24($\pm0.01$) pc.
Most (about 70 percent) of the cores with EYSOs have sizes of $0.15 - 0.4$ pc
with a mean of 0.33 ($\pm 0.03$) pc, while most (more than 85 percent) of starless cores
have sizes of $0.05 - 0.35$ pc with a mean of 0.21 ($\pm 0.01$) pc.
The sizes of cores with EYSOs clearly tend to be larger than those of 
starless cores. Interestingly, this tendency is similar to that of the distribution of the sizes of 
NH$_3$ cores by BM.

We investigated whether this apparent difference in core sizes originates from any distance effect.
Fig. 6 shows the distribution of distances of starless cores and of cores with EYSOs, 
indicating no significant difference in the distances between these two groups.  
Moreover the mean distances are nearly same within
the range of error; about $227\pm8$ pc for starless cores and about $244\pm 19$ pc for cores with EYSOs.
Thus it seems real that the cores with EYSOs are usually 
larger than starless cores by a typical factor of 1.6. 
This may mean that larger cores have a greater likelihood to form stars 
since they have more initial mass, or alternatively cores with EYSOs may be larger 
because they have been growing for a longer time.

The aspect ratio (a/b) of cores  ranges mostly between $1.5\sim 2.5$ for 
both starless cores and cores with EYSOs (Fig. 7).
The mean ($\pm$s.e.m.) aspect ratios are $2.4\pm 0.1$, $2.2\pm0.2$ and $2.4\pm 0.1$
for starless cores, cores with YSOs, and all cores, respectively.
The most elongated cores are L1271-1 (10.4), and L698 (10.0).
Our measured aspect ratios are slightly larger than  
CB' value (2.0) for their cores,
Wood et al. (1994)'s value (1.98) for the IRAS cores, and the value ($1.9\pm 0.12$)
for BM's 41 NH$_3$ cores.
Our measurement for the aspect ratio of cores confirms that
most cores are more likely elongated than spherical. 

{\bf \it 2.2.3 Spatial distribution of selected cores}

Fig. 8 shows the spatial distribution of cores in galactic coordinates. This indicates
that the cores have been uniformly selected from all regions. 
There seems to be no indication that the distribution of cores with EYSOs is different from 
that of starless cores.
In this figure we marked five distinctive star forming regions in boxes that 
contain cores 
with similar distances. They are Taurus
(Ungerechts \& Thaddeus 1987), Ophiuchus (de Geus et al. 1990), Lupus (Murphy et al. 1986), 
Chamaeleon II, III (Chen et al. 1997), and Vela (Vilas-Boas et al. 1994).
The boundaries of each areas were adopted from the references in the parentheses. 

{\bf \it 2.2.4 Implications of the ratio of the number of cores with EYSOs to the number of starless cores}

Our sample gives a ratio $\rm {N(cores~with~EYSOs)\over N(starless~cores)}$ of about 0.31($\approx{94\over 306}$).
This ratio is important in constraining models of core formation,
of which ambipolar diffusion is the best known (Mestel \& Spitzer 1956).
If every core undergoes ambipolar diffusion (i.e, dense core formation), infall, and then 
protostellar emission, the value of the ratio $\rm {N(cores~with~EYSOs) \over
N(starless~cores)}$ should be  approximately  $\rm {t(life~of~protostar)\over t(core~formation)}$,
assuming that the cores have uncorrelated start times.
The time $\rm t(life~of~protostar)$ 
will correspond approximately to the duration of the Class 0 + Class I stages, based on our IRAS
selection criteria described in $\S$2.1. This time interval 
has been observationally determined to be about $(1 - 5) \times 10^5$ years
from the relative numbers of embedded objects and
T Tauri stars (Myers et al. 1987; Wilking et al. 1989; Kenyon et al. 1990).  
The time $\rm t(core~formation)$ should be the time that the core is detectable at its typical observed 
density, according to its formation model.
The cores in our catalogue (most of which have been selected from Lynds classes 5 and 6) 
are thought to have visual extinctions of $4.5 - 6$ magnitudes in their centers  
(Cernicharo \& Bachiller 1984). This means that  the  mean number density in the center of 
our optically selected cores is about $ 6-8\times 10^3$ cm$^{-3}$, assuming
a core diameter of 0.24 pc from our statistics and the relationship  
$\rm N(H_2) \approx 1\times 10^{21}~A_v~cm^{-2}$ from Dickman (1978)
where N(H$_2$) and $\rm A_v$ are column density of H$_2$ and 
magnitude of visual extinction, respectively.
Therefore the time that the core has gas  density $6 - 8\times 10^3$ 
cm$^{-3}$ can be obtained from ambipolar diffusion models. 
We find that different models give slightly different time periods for gas to have 
such density. 
The most detailed model ``$\rm B_{UV}$'' of Ciolek \& Mouschovias (1995) 
gives  about $5.5-6.7\times 10^6$ years, 
the model by Li (1998) about $3.5 - 4.1\times 10^6$ years, and 
the model by Palla \& Galli (1997) about $8.7 - 13.4\times 10^6$ years 
for a core with a size of 0.24 pc and magnetic field strength of 30 $\mu G$. 
Thus present ambipolar diffusion models seem to indicate that detectable cores in our sample
will be condensed within about $3.5 - 13.4 \times 10^6$ years,  after which the cores will undergo 
dynamical, star-forming collapse.
Thereby, the predicted ratios of $\rm {t(life~of~protostar)\over t(core~formation)}$ is
${(1 - 5) \times 10^5 \over (3.5 - 13.4)\times 10^6} \approx 0.007 - 0.14$, which is 
smaller than our empirical value (0.31) by a factor of about $2 - 44$.
This discrepancy between model predictions and our data is also seen for the five star forming regions;
the empirical ratios are 0.39 ($\approx{14\over 36}$) for Taurus, 0.46 ($\approx {19\over 41}$) 
for Ophiuchus, 0.33
($\approx{9\over 27}$) for Lupus, 0.23 ($\approx{3\over 13}$) for 
Chamaeleon II and III, and 0.38 ($\approx {5\over 13}$) for Vela.

As a result, the time scale of
ambipolar diffusion for core formation in the present models seems to be too long,
suggesting that the  models of ambipolar diffusion need to be modified, or
perhaps that faster mechanisms of core formation should be also considered (c.f.,
Tafalla et al. 1998). We suggest that the typical 
lifetime of starless cores ($\ga 6 - 8\times 10^3$ cm$^{-3}$)
may be about 3 times longer than the duration of the Class 0 and Class I phases, i.e.,
about $0.3-1.6\times 10^6$ years, which is shorter than the ambipolar 
model prediction by a factor of $2-44$.
A similar conclusion was reached by Jijina, Myers, \& Adams (1998) based on statistics of 
$\rm NH_3$ cores with and without associated stars.

\section{CONCLUSION}
In this paper we present  a new catalogue of 406 dense cores optically selected 
by using the STScI Digitized Sky Survey (DSS).
We have imaged areas of 0.5 or 1.0 square degree around already known 
numerous NH$_3$ and C$^{18}$O cores, every Lynds class 5 and 6 clouds (Lynds 1962),
southern Hartley et al. (1986)' class A clouds, to identify every local minimum of intensity
in each field, and to select the darkest and best resolved minima. 
For the identified 406 sources, we used the IRAS Point Source
Catalogue to determine which have associated point sources, and used 
the criteria for PMS star by Weintraub (1990) and the catalogue of PMS star by Herbig \& Bell (1988)
to find possible PMS stars associated with the cores.
From these surveys, 306 starless cores without  an Embedded YSO (EYSO)
nor a Pre-Main-Sequence (PMS) star, 94 cores with EYSOs (1 core has  both an EYSO and a PMS star),
and 6 cores with PMS star only are found.  
This list of dense cores will be a useful database for an extensive  systematic study of dense cores
for a better understanding of the early stages of star formation.

Our fairly complete sample of most opaque cores gives representative statistics of physical 
parameters of dense cores.
Most cores are found to have an apparent angular size of $1'\sim4'$ and 
a linear size of $\sim 0.1-0.3$ pc, with an average of 0.24 pc.
Interestingly, the sizes (mean: $0.33 \pm 0.03$ pc) of the cores with EYSOs are usually larger than 
those (mean: $0.21 \pm 0.01$ pc) of the starless cores.
The mean aspect ratio of cores is $2.4\pm 0.1$,  confirming that most cores are more likely 
elongated than spherical.

The ratio of $\rm {N(cores~with~EYSOs)\over N(starless~cores)}$ for our sample 
is about 0.31($\approx{94\over 306}$).
This value is much higher than expected 
from the relative time scales (${t(life~of~protostar)\over t(core~formation)}
\approx {(1 - 5) \times 10^5 \over (3.5 - 13.4)\times 10^6} \approx 0.007 - 0.14$)
of ambipolar diffusion of a core (with gas density $\ga 6 - 8\times 10^3$ cm$^{-3}$) 
and the life of a protostar.
The statistics of our sample indicate that the ambipolar diffusion time scale predicted by
present models  for dense core formation may be too long.
The typical lifetime of starless cores is suggested to be about 3 times longer
than  the duration of the Class 0 and Class I phases, i.e.,
about $0.3-1.6\times 10^6$ years, which is shorter than the ambipolar
model predictions by a factor of $2-44$.
 
\acknowledgments
C.W.L acknowledges the financial support by 1996 Overseas Postdoctoral Support Program of 
Korea Research Foundation. C.W.L also thanks the Harvard-Smithsonian Center for Astrophysics 
for support while working on this project. We would like to thank Lori Allen and 
August Muench for their careful reading of our manuscript and an anonymous referee for his
useful comments.
This research was  supported  by NASA Origins of 
Solar System Program, Grant NAGW-3401.
\vfill\eject

\clearpage
\begin{figure}
\begin{center}
{\bf FIGURE CAPTIONS}
\end{center}
%\plotone {fig1.ps}
\caption{
An example (for L1622) of how to select the local minimum
positions of the intensity in a cloud by its optical extinction 
using an image extracted from the STScI Digitized Sky Survey.
The area of the image is $0^\circ.5\times 0^\circ.5$ and the coordinate of the center is
($\alpha, \delta)_{1950.0}=(5^h~52^m~ 00^s.0,~01^\circ~51'~00'')$,
which was changed from Lynds (1962) position to shift the
cloud into a central position in the image.
The smooth contours are overlaid on the gray scale image to enable identification of
two local minima (A and B) of intensity.
The smoothing length for the contours in this case is $30''$.
The regions with the locally lowest or highest intensity are marked with
characters `L' and `H', respectively. The numeric values below the characters
are data values by DSS that are proportional to photographic densities.}
%Thick contours are FWHMs for the selected cores A \& B. 
\label{l1622.sample}
\end{figure}

\begin{figure}
\caption{An example of a cut-profile of L1622 cloud. This cut-profile was obtained 
by crossing two intensity minima of L1622A and L1622B.
The angular size for 1 pixel in abscissa is about $0.6''$. 
The data value (DN) is on a scale of photographic density ($\gamma$) in 
the expression of $\gamma = DN/6553.4$ (Postman 1996).
The data values ($I_B$ and $I_C$) of DSS for the background and the core
are approximately adopted from this profile.}
\label{cu-profile}
\end{figure}

\begin{figure}
\caption{Relation of the Darkness Contrast (DC) to Lynds class 5, 6 and HMSTG class A.
Lynds 6 and HMSTG A clouds more likely have a DC of 3 or 4 than Lynds 5 clouds.}
\label{con-lynds}
\end{figure}

\begin{figure}
%\plotone {fig4.ps}
\caption{ Distribution of the apparent angular size of cores. The positions of
mean angular sizes (mean$\pm$s.e.m.) of  $3.7'\pm0.1'$,
$3.3'\pm0.1'$, and $5.1'\pm0.4'$ for all cores, starless cores, and cores with EYSOs, 
respectively, are marked with vertical bars in each panel.}
\label{asize}
\end{figure}

\begin{figure}
%\plotone {fig5.ps}
\caption{ Distribution of the linear size of cores. The positions of 
mean linear sizes (mean$\pm$s.e.m.) of  $0.24\pm0.01$ pc,
$0.22\pm0.01$ pc, and $0.33\pm0.02$ pc for all cores, starless cores, and cores with EYSOs, 
respectively, are marked with vertical bars in each panel.}
\label{asize}
\end{figure}

\begin{figure}
%\plotone {fig6.ps}
\caption{ Distance distribution of cores.
The distributions of distances of starless cores and cores with EYSO  are not different.
The positions of mean distance (mean$\pm$s.e.m.) of $231\pm8$ pc,
$227\pm8$ pc, and $244\pm19$ pc for all cores, starless cores, and cores with EYSOs,
respectively, are indicated with vertical bars in each panel.}
\label{dist_cores}
\end{figure}

\begin{figure}
%\plotone {fig7.ps}
\caption{ Distribution of the aspect ratio for all cores, starless cores,
and cores with EYSOs. The positions of the mean aspect ratio (mean$\pm$s.e.m.) of $2.4\pm0.1$,
$2.4\pm0.1$, and $2.2\pm0.2$ for all cores, starless cores, and cores with EYSOs, 
respectively, are indicated with vertical bars in each panel.}
\label{aspect}
\end{figure}

\begin{figure}
%\plotone {fig8.ps}
\caption{Distributions in galactic coordinates of (a) cores with EYSOs and (b) starless cores 
listed in Table 2.}
\label{distri_lb}
\end{figure}

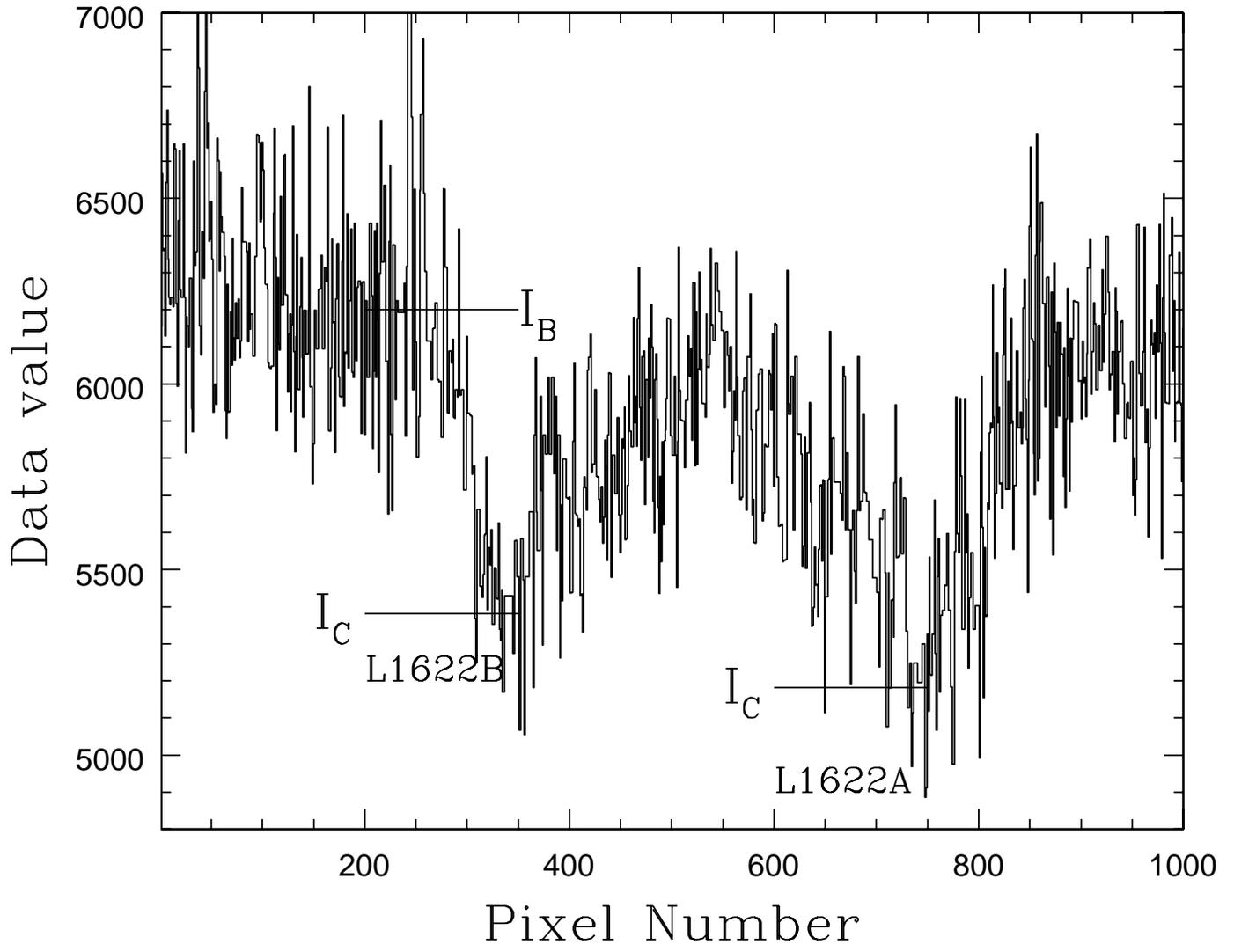
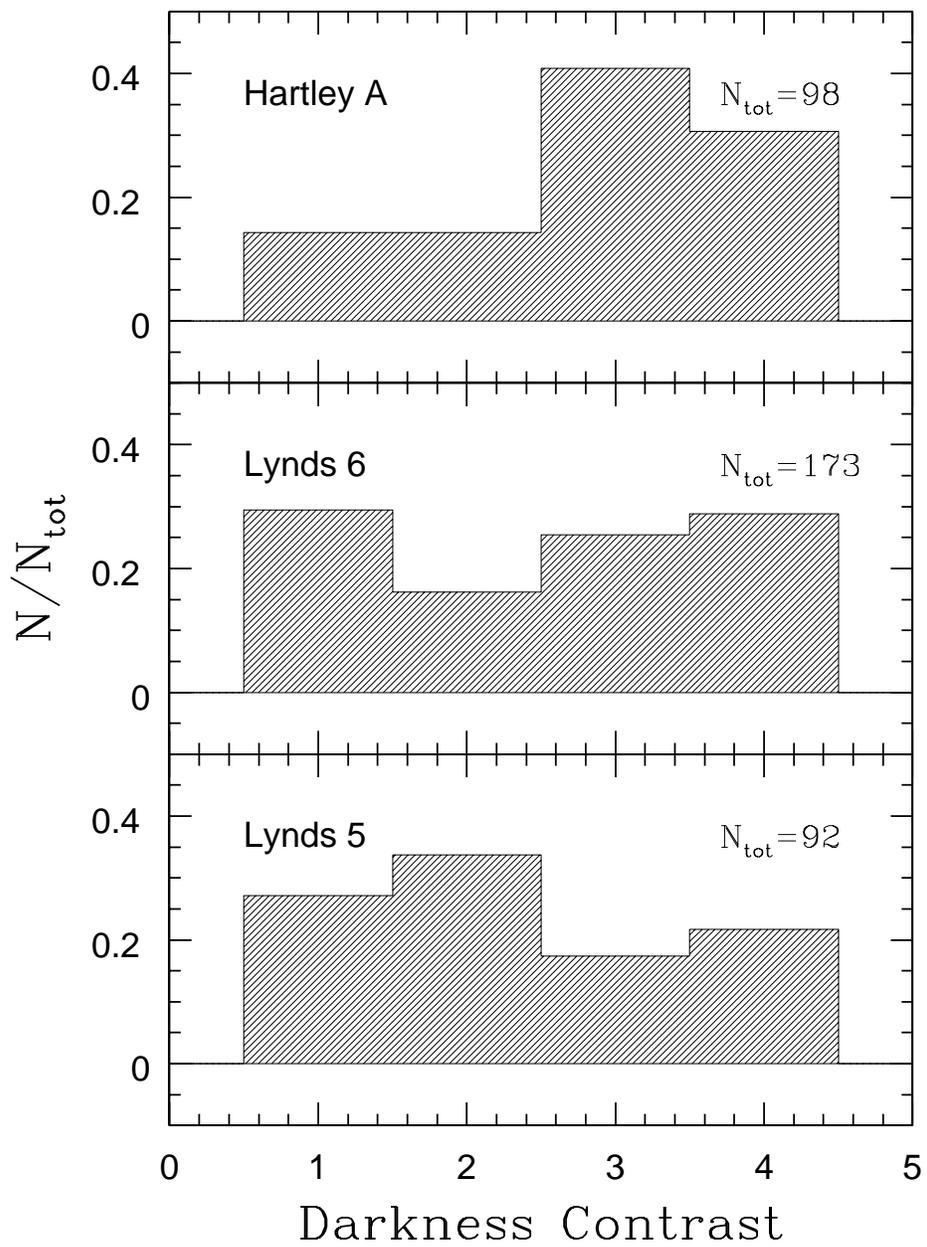
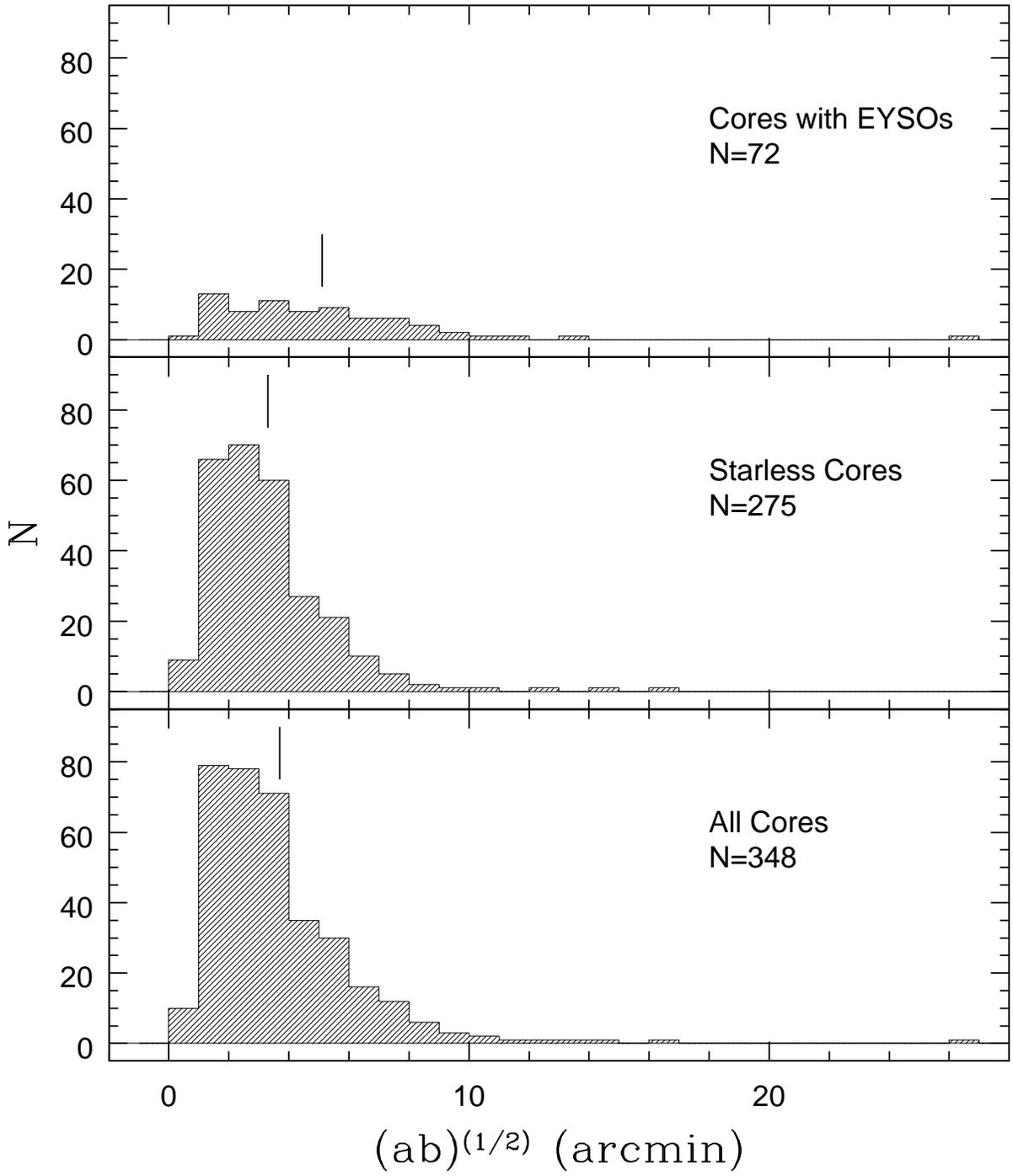
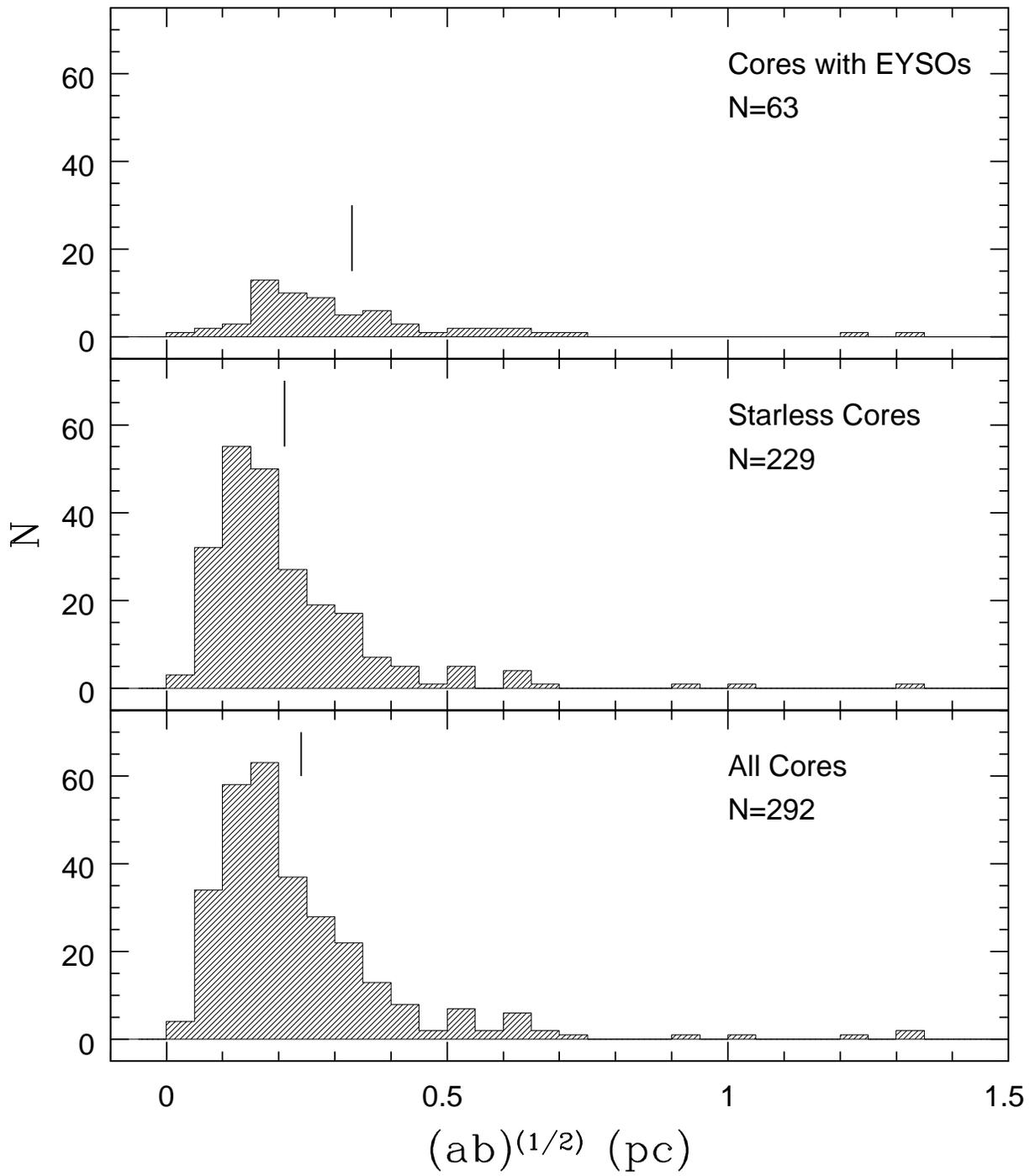
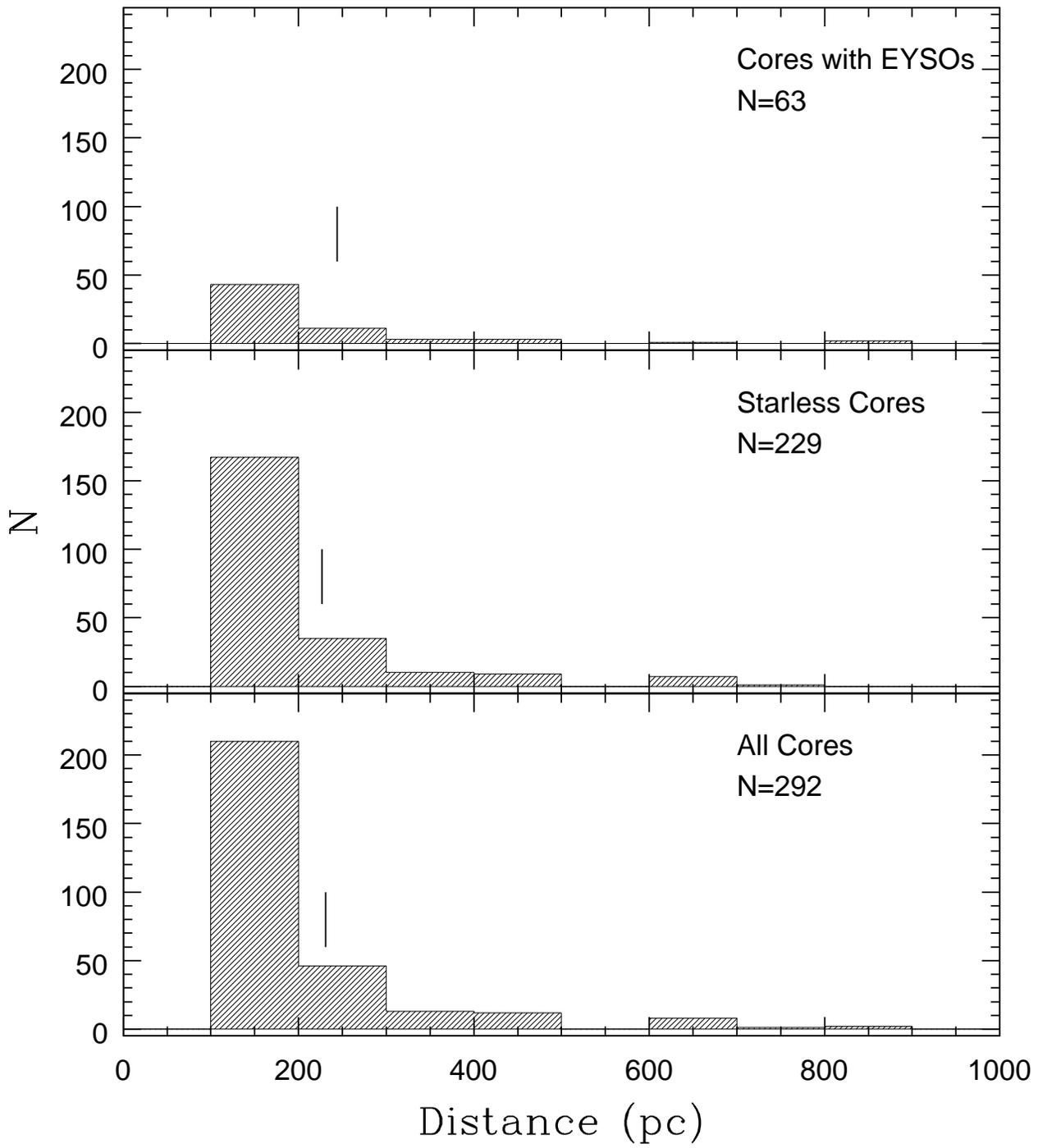
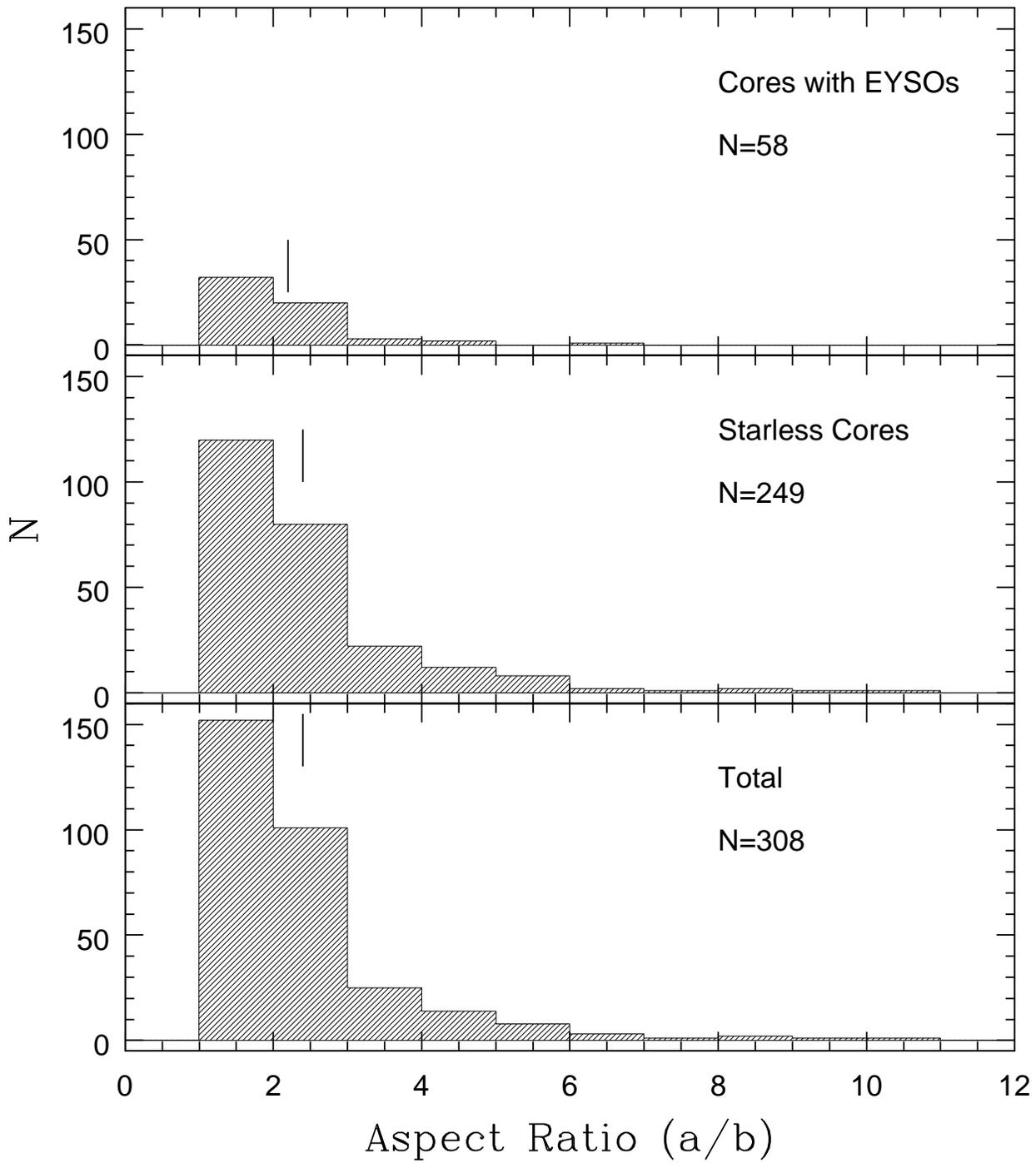
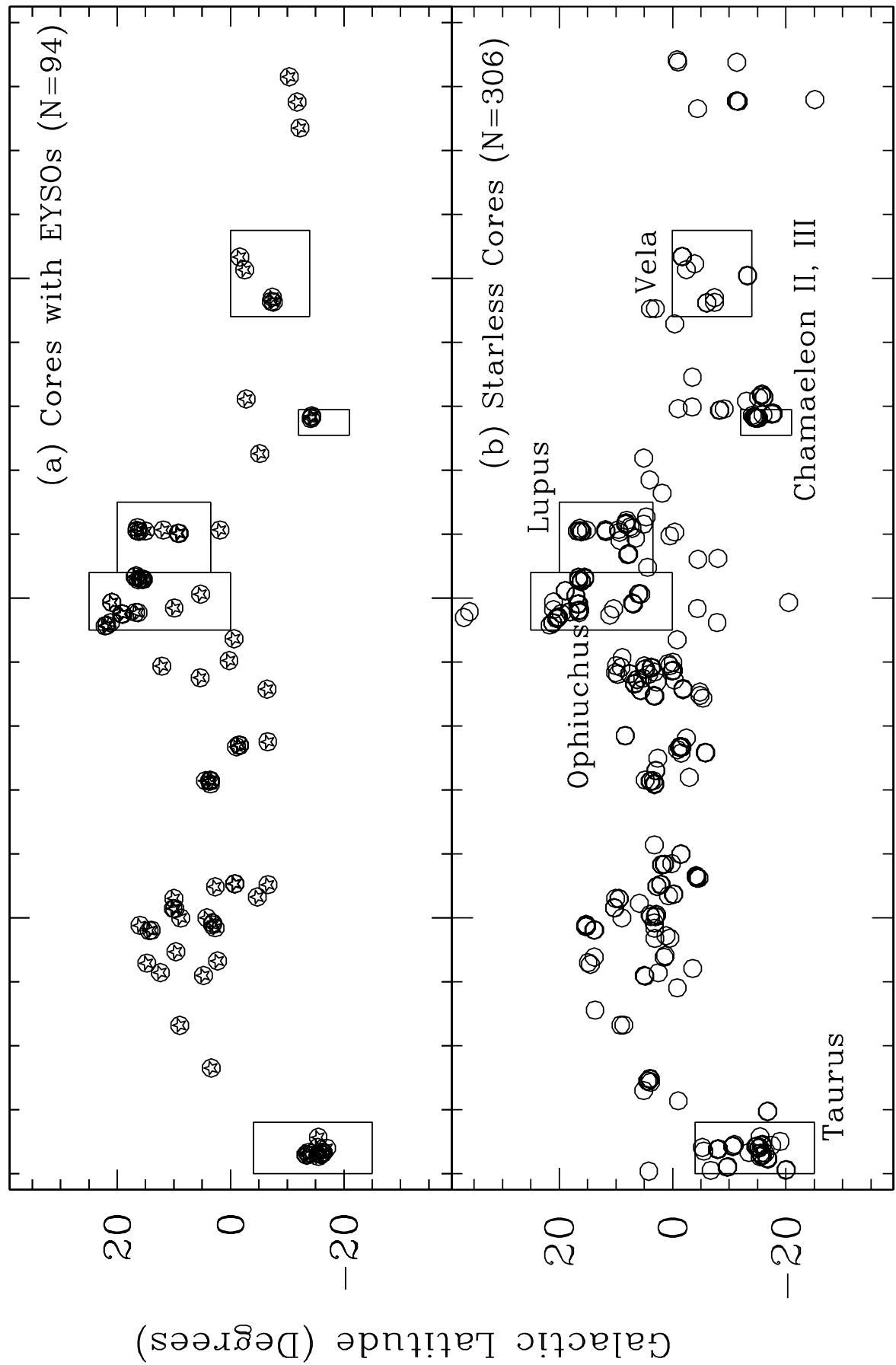
\end{document}

